\documentstyle[epsf,epsfig,here,12pt]{article}

\newcommand{\vcb}{|V_{cb}|}
\newcommand{\vtd}{|V_{td}|}
\newcommand{\vub}{|V_{ub}/V_{cb}|}
\newcommand{\vts}{|V_{ts}|}
\newcommand{\vus}{|V_{us}|}

\newcommand{\bea}{\begin{eqnarray}}
\newcommand{\eea}{\end{eqnarray}}
\newcommand{\bd}{\begin{displaymath}}
\newcommand{\ed}{\end{displaymath}}
\newcommand{\beq}{\begin{equation}}
\newcommand{\eeq}{\end{equation}}
\newcommand{\be}{\begin{equation}}
\newcommand{\ee}{\end{equation}}
\newcommand{\bi}{\begin{itemize}}
\newcommand{\ei}{\end{itemize}}
\newcommand{\ord}{{\cal O}}

\begin{document}
\thispagestyle{empty}
\phantom{xxx}
\begin{flushright}
  LAL 02-39 \\
 TUM-HEP-465/02\\
 October 2002
\end{flushright}
\vskip1.5truecm
\centerline{\LARGE\bf The CKM Matrix and The Unitarity Triangle:}
\centerline{\LARGE\bf Another Look}
   \vskip1truecm
\centerline{\Large\bf Andrzej J. Buras${}^{1)}$, 
Fabrizio Parodi${}^{2)}$ and 
Achille Stocchi${}^{3)}$}
\bigskip
\centerline{\sl  
${}^{1)}$ Technische Universit{\"a}t M{\"u}nchen, Physik Department}
\centerline{\sl D-85748 Garching, Germany}

\centerline{\sl ${}^{2)}$  Dipartimento di Fisica, 
Universita' di Genov{a} and INFN}
\centerline{\sl Via Dodecaneso 33, 16146 Genova, Italy}

\centerline{\sl  ${}^{3)}$ Laboratoire de l'Acc{\'e}l{\'e}rateur 
Lin{\'e}aire}
\centerline{\sl IN2P3-CNRS et Universit\'e de Paris-Sud, BP 34, F-91898 Orsay Cedex}
\centerline{\sl and CERN, 1211 Geneva 23, Switzerland}

\vskip 0.8truecm
\centerline{\bf Abstract}
The unitarity triangle can be determined by means of two measurements 
of its sides or angles. Assuming the same relative errors on the
angles $(\alpha,\beta,\gamma)$ and the sides $(R_b,R_t)$, we find that the
pairs $(\gamma,\beta)$ and $(\gamma,R_b)$ are most efficient in determining
$(\bar\varrho,\bar\eta)$ that describe the apex of the unitarity triangle. 
They are followed by  
$(\alpha,\beta)$, $(\alpha,R_b)$, $(R_t,\beta)$, $(R_t,R_b)$ and 
$(R_b,\beta)$.
As the set $\vus$, $\vcb$, $R_t$ and $\beta$ appears to be the best 
candidate for the fundamental set of flavour violating parameters in
the coming years, we show various constraints on the CKM matrix in the
$(R_t,\beta)$ plane. Using the best available input we determine the
universal unitarity triangle for models with minimal flavour violation
(MFV) and compare it with the one in the Standard Model. We present allowed
ranges for $\sin 2\beta$, $\sin 2\alpha$, $\gamma$, $R_b$, $R_t$ and 
$\Delta M_s$ 
within the Standard Model and MFV models. We also 
update the allowed range for the function $F_{tt}$ that parametrizes
various MFV-models.

\newpage

\section{Introduction}
\label{sec:intro}
\setcounter{equation}{0}
The determination of the Cabibbo-Kobayashi-Maskawa (CKM) matrix \cite{CAB,KM}
that parametrizes the weak charged current interactions of quarks is one of
the important targets of particle physics. During the last two decades
several strategies have been proposed that should allow one to determine 
the CKM matrix and  the related unitarity triangle (UT). They 
are reviewed in particular in \cite{BABAR,LHCB,FERMILAB,BF97,Erice}.

To be specific let us first choose as the independent parameters
\begin{equation}\label{I1}
\vus, \qquad \vcb, \qquad \bar\varrho, \qquad \bar\eta
\end{equation}
where $(\bar\varrho,\bar\eta)$, defined below, determine the apex of 
the unitarity triangle in question.
The best place to determine $\vus$ and $\vcb$ are the semi-leptonic K and 
B decays, respectively. The question that we want address here is the
determination of the remaining two parameters $(\bar\varrho,\bar\eta)$.

There are many ways to determine $(\bar\varrho,\bar\eta)$. As the length 
of  one side
of the rescaled unitarity triangle is fixed to unity, we have to our disposal
two sides, $R_b$ and $R_t$ and three angles, $\alpha$, $\beta$ and $\gamma$.
These five quantities can be measured by means of rare K and B 
decays and in particular by studying CP-violating observables. While until
recently only a handful of strategies could be realized, the present decade
should allow several independent determinations of $(\bar\varrho,\bar\eta)$ 
that will test the KM picture of CP violation \cite{KM} and possibly indicate
the physics beyond the Standard Model (SM).

The determination of $(\bar\varrho,\bar\eta)$ in a given strategy is subject 
to experimental and theoretical errors and it is important to identify 
those strategies that are experimentally feasible and in which
hadronic uncertainties are as much as possible under control. 
Such strategies are reviewed in \cite{BABAR,LHCB,FERMILAB,BF97,Erice}.

Here we want to address a different question. The determination of 
$(\bar\varrho,\bar\eta)$ requires at least two independent measurements. 
In most cases these
are the measurements of two sides of the UT, of one side and one angle or
the measurements of two angles. Sometimes $\bar\eta$ can be directly 
measured and combining it with the knowledge of one angle or one side 
of the UT, $\bar\varrho$ can be found. Analogous comments apply to 
measurements in which $\bar\varrho$ is directly measured. Finally
in more complicated strategies one measures various linear combinations 
of angles, sides or $\bar\varrho$ and $\bar\eta$.

Restricting first our attention to measurements in which sides and angles 
of the UT can be measured independently of each other, we end up with 
ten different pairs of measurements that allow the determination of
$(\bar\varrho,\bar\eta)$. The question then arises which of the pairs 
in question is most efficient in the determination of the UT?
That is, given the same relative errors on $R_b$, $R_t$, $\alpha$, 
$\beta$ and $\gamma$, we want to find which of the pairs gives the most 
accurate determination of $(\bar\varrho,\bar\eta)$. This is one of the 
 questions that we want to address here.

 The answer to this question depends necessarily on
the values of $R_b$, $R_t$, $\alpha$, $\beta$ and $\gamma$ but as we will see
below just the requirement of the consistency of $R_b$ with the measured
value of $|V_{ub}/V_{cb}| $ implies
a hierarchy within the ten strategies mentioned above.

During the 1970's and 1980's the variables $\alpha_{QED}$, the Fermi 
constant $G_F$ and the sine of the Weinberg angle ($\sin\theta_W$) were 
the fundamental parameters in terms of which the electroweak tests of 
the SM have been performed. After the $Z^0$ boson has been discovered
and its mass precisely measured at LEP-I, $\sin\theta_W$ has been replaced
by $M_Z$ and the fundamental set used in the electroweak precision studies 
in the 1990's has been $(\alpha_{QED},G_F,M_Z)$. It is to be expected that
when $M_W$ will be measured precisely this set will be changed to 
$(\alpha_{QED},M_W,M_Z)$ or ($G_F,M_W,M_Z)$.

We anticipate that an analogous development will happen in this decade 
in connection with the CKM matrix. While the set (\ref{I1}) has clearly 
many virtues and has been used extensively in the literature, one should
emphasize that presently no direct independent measurements of $\bar\eta$
and $\bar\varrho$ are available. $|\bar\eta|$ can be measured cleanly in
the decay $K_L\to\pi^0\nu\bar\nu$. On the other hand to our knowledge
there does not exist any strategy for a clean independent measurement 
of $\bar\varrho$. 

Taking into account the experimental feasibility of various measurements
and their theoretical cleanness, the most obvious candidate for the 
fundamental set 
in the quark flavour physics for the coming years 
appears to be
\begin{equation}\label{I2}
\vus, \qquad \vcb, \qquad R_t, \qquad \beta
\end{equation}
with the last two variables describing the $V_{td}$ coupling that can 
be measured by means of the $B^0-\bar B^0$ 
mixing ratio $\Delta M_d/\Delta M_s$ and the CP-asymmetry $a_{\psi K_S}$, 
respectively.
In this context we investigate, in analogy to the $(\bar\varrho,\bar\eta)$ 
plane and the planes $(\sin 2\beta,\sin 2\alpha)$ \cite{betaalpha} and 
$(\gamma,\sin 2\beta)$ \cite{Lacker}
considered in the past, the $(R_t,\beta)$
plane for the exhibition of various constraints on the CKM matrix. 
We also provide the parametrization of the CKM matrix given directly in
terms of the variables (\ref{I2}).

Several of the results and formulae presented here are not entirely new 
and have been already discussed by us and other authors in the 
past. In particular in \cite{BBSIN} it has been pointed out that only a 
moderately precise measurement of $\sin 2\alpha$ can be as useful for 
the UT as a precise measurement of the angle $\beta$. This has been recently 
reemphasized in \cite{BBNS2}. Similarly the measurement of the pair 
$(\alpha,\beta)$ has been found to be a very efficient tool for
the determination of the UT \cite{BLO,B95} and the construction of 
the full CKM matrix from the angles of various unitarity triangles has 
been presented in \cite{Kayser}. Finally the importance of 
the pair $(R_t,\sin 2\beta)$ has been emphasized recently in a number of 
papers
\cite{UUT,ALILOND,BF01,BCRS1,AI01}. 
Many useful relations relevant for the unitarity triangle
 can also be found in \cite{Branco,BOBRNERE}.
On the other hand, to our knowledge,
no systematic classification of the strategies in question and their 
comparison has been presented in the literature and the discussion of
the $(R_t,\beta)$ plane is presented for the first time.
 We think that in view
of the present and  future efforts to determine the CKM matrix such
a study is desirable.

Our paper is organized as follows. In section \ref{sec:ckmmatrix} we recall 
some formulae related to the CKM matrix and the UT. 
In section \ref{sec:strategy} 
we list the expressions 
for $\bar\varrho$ and $\bar\eta$ in the ten strategies in question and 
provide the parametrization of the CKM matrix directly in terms of 
the set (\ref{I2}).
In section \ref{sec:Hierarchies} we present a numerical analysis that 
reveals a hierarchy of various 
determinations and we show how various constraints appear in the different
planes corresponding to the leading strategies.
In section \ref{sec:present} we show the implications of
 the presently available strategies $(R_t,\beta)$, $(R_b,\beta)$ and 
$(R_t,R_b)$ in determining the allowed region for $(\bar\varrho,\bar\eta)$ 
and we present the available constraints on the CKM matrix in the $(R_t,\beta)$ 
plane.
In section \ref{sec:mfvm} we determine the 
universal unitarity triangle for models with minimal flavour violation
(MFV) and compare it with the one in the Standard Model. We also 
update the allowed range for the function $F_{tt}$ that parametrizes
various MFV-models.
We conclude in section \ref{sec:conclusions}.
\section{CKM Matrix and the Unitarity Triangle}
\label{sec:ckmmatrix}
\setcounter{equation}{0}
Many parametrizations of the CKM
matrix have been proposed in the literature.   The most popular are
the standard parametrization 
\cite{CHAU} recommended by the Particle Data Group  \cite{PDG}  
and a generalization of the Wolfenstein parametrization \cite{WO} as 
presented in \cite{BLO}. 

With
$c_{ij}=\cos\theta_{ij}$ and $s_{ij}=\sin\theta_{ij}$ 
($i,j=1,2,3$), the standard parametrization is
given by:
\begin{equation}\label{2.72}
\hat V_{\rm CKM}=
\left(\begin{array}{ccc}
c_{12}c_{13}&s_{12}c_{13}&s_{13}e^{-i\delta}\\ -s_{12}c_{23}
-c_{12}s_{23}s_{13}e^{i\delta}&c_{12}c_{23}-s_{12}s_{23}s_{13}e^{i\delta}&
s_{23}c_{13}\\ s_{12}s_{23}-c_{12}c_{23}s_{13}e^{i\delta}&-s_{23}c_{12}
-s_{12}c_{23}s_{13}e^{i\delta}&c_{23}c_{13}
\end{array}\right)\,,
\end{equation}
where $\delta$ is the phase necessary for {\rm CP} violation.
$c_{ij}$ and
$s_{ij}$ can all be chosen to be positive
and  $\delta$ may vary in the
range $0\le\delta\le 2\pi$. However, the measurements
of CP violation in $K$ decays force $\delta$ to be in the range
 $0<\delta<\pi$. 

>From phenomenological applications we know that 
$s_{13}$ and $s_{23}$ are small numbers: $\ord(10^{-3})$ and ${\cal
O}(10^{-2})$,
respectively. Consequently, to an excellent accuracy, the four 
independent parameters are given as 
\begin{equation}\label{2.73}
s_{12}=| V_{us}|, \quad s_{13}=| V_{ub}|, \quad s_{23}=|
V_{cb}|, \quad \delta.
\end{equation}

The first three can be extracted from tree level decays mediated
by the transitions $s \to u$, $b \to u$ and $b \to c$ respectively.
The phase $\delta$ can be extracted from CP violating transitions or 
loop processes sensitive to $\vtd$.

For our purposes it will be convenient to make the following
change of variables in (\ref{2.72}) \cite{BLO,schubert}
\begin{equation}\label{2.77} 
s_{12}=\lambda\,,
\qquad
s_{23}=A \lambda^2\,,
\qquad
s_{13} e^{-i\delta}=A \lambda^3 (\varrho-i \eta)
\end{equation}
where $\lambda$, $A$, $\varrho$ and $\eta$ are Wolfenstein parameters. 
It follows then that to a very good accuracy the set in (\ref{2.73})
can be replaced by
\begin{equation}\label{I1a}
\vus=\lambda, \qquad \vcb, \qquad \bar\varrho, \qquad \bar\eta
\end{equation}
where \cite{BLO}
\begin{equation}\label{2.88d}
\bar\varrho=\varrho (1-\frac{\lambda^2}{2}),
\qquad
\bar\eta=\eta (1-\frac{\lambda^2}{2}).
\end{equation}

\begin{figure}[hbt!]
\begin{center}
{\epsfig{figure=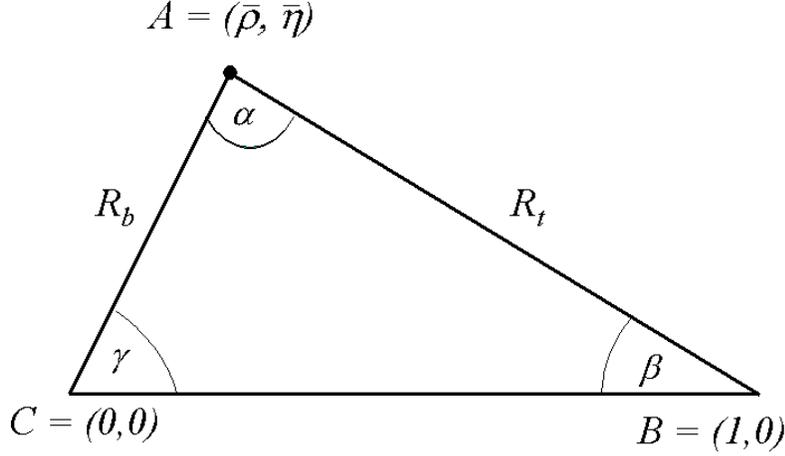,height=8cm}}
\caption{Unitarity Triangle.}
\label{fig:utria}
\end{center}
\end{figure}

The pair $(\bar\varrho,\bar\eta)$ describes the apex of the unitarity 
triangle shown in figure \ref{fig:utria} that represents 
the unitarity relation
\begin{equation}\label{2.87h}
V_{ud}^{}V_{ub}^* + V_{cd}^{}V_{cb}^* + V_{td}^{}V_{tb}^* =0
\end{equation}
suitably rescaled by $| V_{cd}^{}V_{cb}^*|=A\lambda^3=\lambda\vcb$, 
with the latter equalities satisfied to an excellent accuracy \cite{Erice,BLO}.

Let us collect useful formulae related to this triangle:
\bi
\item
The lengths $CA$ and $BA$ to be denoted by $R_b$ and $R_t$,
respectively, are given by
\begin{equation}\label{2.94}
R_b \equiv \frac{| V_{ud}^{}V^*_{ub}|}{| V_{cd}^{}V^*_{cb}|}
= \sqrt{\bar\varrho^2 +\bar\eta^2}
= (1-\frac{\lambda^2}{2})\frac{1}{\lambda}
\left| \frac{V_{ub}}{V_{cb}} \right|,
\end{equation}
\begin{equation}\label{2.95}
R_t \equiv \frac{| V_{td}^{}V^*_{tb}|}{| V_{cd}^{}V^*_{cb}|} =
 \sqrt{(1-\bar\varrho)^2 +\bar\eta^2}
=\frac{1}{\lambda} \left| \frac{V_{td}}{V_{cb}} \right|.
\end{equation}
\item
The angles $\beta$ and $\gamma=\delta$ of the unitarity triangle are related
directly to the complex phases of the CKM-elements $V_{td}$ and
$V_{ub}$, respectively, through
\beq\label{e417}
V_{td}=|V_{td}|e^{-i\beta},\qquad V_{ub}=|V_{ub}|e^{-i\gamma}.
\eeq
\item
The unitarity relation (\ref{2.87h}) can be rewritten as
\be\label{RbRt}
R_b e^{i\gamma} +R_t e^{-i\beta}=1~.
\ee
\item
The angle $\alpha$ can be obtained through the relation
\beq\label{e419}
\alpha+\beta+\gamma=180^\circ
\eeq
expressing the unitarity of the CKM-matrix.
\ei

The triangle depicted in figure \ref{fig:utria},  $|V_{us}|$ 
and $\vcb$ give the full description of the CKM matrix. 

Formula (\ref{RbRt}) shows transparently that the knowledge of
$(R_t,\beta)$ allows to determine $(R_b,\gamma)$ through \cite{BCRS1}
\be\label{VUBG}
R_b=\sqrt{1+R_t^2-2 R_t\cos\beta},\qquad
\cot\gamma=\frac{1-R_t\cos\beta}{R_t\sin\beta}.
\ee
Similarly, $(R_t,\beta)$ can be expressed through $(R_b,\gamma)$:
\be\label{VTDG}
R_t=\sqrt{1+R_b^2-2 R_b\cos\gamma},\qquad
\cot\beta=\frac{1-R_b\cos\gamma}{R_b\sin\gamma}.
\ee
These formulae relate the leading strategy $(R_t,\beta)$ for the 
determination of the so-called {\it universal unitarity triangle} 
\cite{UUT} within the models with minimal flavour violation (MFV)
\cite{CDGG} and the strategy 
$(R_b,\gamma)$ that results in the so-called {\it reference unitarity 
triangle} as proposed and discussed in \cite{refut}.

\section{General Strategies}
\label{sec:strategy}
\setcounter{equation}{0}
\subsection{Basic Formulae}
We list below the formulae for $\bar\varrho$ and $\bar\eta$ in 
the strategies that are labelled by the two measured quantities as
discussed at the beginning of our paper.  
\subsubsection{\boldmath{$R_t$} and \boldmath{$\beta$}} 
\be\label{S1}
\bar\varrho=1-R_t\cos\beta,\qquad \bar\eta=R_t\sin\beta~.
\ee

\subsubsection{\boldmath{$R_b$} and \boldmath{$\gamma$}} 
\be\label{S2}
\bar\varrho=R_b\cos\gamma,\qquad \bar\eta=R_b\sin\gamma~.
\ee

\subsubsection{\boldmath{$R_b$} and \boldmath{$R_t$}} 
\be\label{S3}
\bar\varrho = \frac{1}{2} (1+R^2_b-R^2_t),
\qquad \bar\eta=\sqrt{R_b^2-\bar\varrho^2}
\ee
where $\bar\eta>0$ has been assumed.

\subsubsection{\boldmath{$R_t$} and \boldmath{$\gamma$}} 
This strategy uses (\ref{S2}) with
\be\label{S4}
R_b=\cos\gamma\pm \sqrt{R^2_t-\sin^2\gamma}~.
\ee
The two possibilities can be distinguished by the measured 
value of $R_b$.

\subsubsection{\boldmath{$R_b$} and \boldmath{$\beta$}} 
This strategy uses (\ref{S1}) and
\be\label{S5}
R_t=\cos\beta\pm \sqrt{R^2_b-\sin^2\beta}~.
\ee
The two possibilities can be distinguished by the measured 
value of $R_t$.

\subsubsection{\boldmath{$R_t$} and \boldmath{$\alpha$}} 
\be\label{S6a}
\bar\varrho=1-R_t^2\sin^2\alpha+R_t\cos\alpha\sqrt{1-R_t^2\sin^2\alpha},
\ee
\be\label{S6b}
\bar\eta=R_t\sin\alpha\left[R_t\cos\alpha+\sqrt{1-R_t^2\sin^2\alpha}\right] 
\ee
where $\cos\gamma>0$ has been assumed. For $\cos\gamma<0$ the signs in front
of the square roots should be reversed.

\subsubsection{\boldmath{$R_b$} and \boldmath{$\alpha$}} 
\be\label{S7a}
\bar\varrho=R_b^2\sin^2\alpha-R_b\cos\alpha\sqrt{1-R_b^2\sin^2\alpha},
\ee
\be\label{S7b}
\bar\eta=R_b\sin\alpha\left[R_b\cos\alpha+\sqrt{1-R_b^2\sin^2\alpha}\right]
\ee
where $\cos\beta>0$ has been assumed.

\subsubsection{\boldmath{$\beta$} and \boldmath{$\gamma$}} 
\be\label{S8}
R_t=\frac{\sin\gamma}{\sin(\beta+\gamma)},
\qquad R_b=\frac{\sin\beta}{\sin(\beta+\gamma)}
\ee
and (\ref{S3}).

\subsubsection{\boldmath{$\alpha$} and \boldmath{$\gamma$}} 
\be\label{S9}
R_t=\frac{\sin\gamma}{\sin\alpha},
\qquad R_b=\frac{\sin(\alpha+\gamma)}{\sin\alpha}
\ee
and (\ref{S3}).

\subsubsection{\boldmath{$\alpha$} and \boldmath{$\beta$}} 
\be\label{S10}
R_t=\frac{\sin(\alpha+\beta)}{\sin\alpha},
\qquad R_b=\frac{\sin\beta}{\sin\alpha}
\ee
and (\ref{S3}).

Finally we give the formulae for the strategies in which $\bar\eta$ is 
directly measured and the strategy allows to determine $\bar\varrho$.
\subsubsection{\boldmath{$\bar\eta$} and \boldmath{$R_t$} or \boldmath{$R_b$}} 
\be\label{S11}
\bar\varrho=1-\sqrt{R_t^2-\bar\eta^2}, \qquad
\bar\varrho=\pm\sqrt{R_b^2-\bar\eta^2}
\ee
where in the first case we have excluded the + solution in view of 
 $R_b\le 0.5$ as extracted from the experimental data on $\vub$. 
\subsubsection{\boldmath{$\bar\eta$} and \boldmath{$\beta$} or \boldmath{$\gamma$}}
\be\label{S12}
\bar\varrho=1-\frac{\bar\eta}{\tan\beta}, \qquad
\bar\varrho=\frac{\bar\eta}{\tan\gamma}~.
\ee

\subsection{CKM Matrix and the Fundametal Variables}
It is useful for phenomenological purposes to express the CKM matrix 
directly in terms of the parameters selected in a given strategy.
This can be easily done by inserting the formulae for $\bar\varrho$ 
and $\bar\eta$ presented here into the known expressions for 
the CKM elements in terms of these variables \cite{WO,BLO} as given 
in section \ref{sec:ckmmatrix}.

Here we give explicit result only for the set (\ref{I2}). In order to 
simplify the notation we use $\lambda$ instead of $\vus$ as
$V_{us}=\lambda+\ord(\lambda^7)$. We find 

\be
V_{ud}=1-\frac{1}{2}\lambda^2-\frac{1}{8}\lambda^4 +\ord(\lambda^6),
\qquad
V_{ub}=\frac{\lambda}{1-\lambda^2/2}\vcb \left[1-R_t e^{i\beta}\right],
\ee
\be
V_{cd}=-\lambda+\frac{1}{2} \lambda \vcb^2 -
\lambda\vcb^2 \left[1-R_t e^{-i\beta}\right] +
\ord(\lambda^7),
\ee
\be
V_{cs}= 1-\frac{1}{2}\lambda^2-\frac{1}{8}\lambda^4 -\frac{1}{2} \vcb^2
 +\ord(\lambda^6),
\ee
\be
V_{tb}=1-\frac{1}{2} \vcb^2+\ord(\lambda^6),
\qquad
V_{td}=\lambda\vcb R_t e^{-i\beta}
+\ord (\lambda^7),
\ee
\begin{equation}\label{2.83d}
 V_{ts}= -\vcb +\frac{1}{2} \lambda^2 \vcb - 
\lambda^2 \vcb \left[1-R_t e^{-i\beta}\right]
  +\ord(\lambda^6)~.
\end{equation}

\section{Hierarchies}
\label{sec:Hierarchies}
\setcounter{equation}{0}
The numerical analysis  of various strategies listed in the previous section was performed using
a Bayesian approach as described in \cite{ref:haricot}.\\
Considering two measured quantities $x$ and $y$ the bidimensional probability density function 
for $x$ and $y$ is given (after appropriate normalization) by:
\begin{eqnarray*}
  f(x, y) \propto {\cal L}(x) {\cal L}(y) f_o(x,y)
\end{eqnarray*}
where $f_o$ is the prior probability density function for $x$, $y$ (taken uniform all over 
the range) 
and ${\cal L}(x)$, ${\cal L}(y)$ represent the likelihood functions of the two 
measurements (assumed as a Gaussian distribution).\\
The probability density function for $\bar \eta$ and $\bar \rho$ ($f(\bar \rho, \bar \eta)$)
is then obtained from $f(x,y)$ by changing the variables using the equations of the previous 
chapter. The probability density functions for  $\bar \rho$ ($\bar \eta$) 
is then obtained by integrating $f$ over $\bar \eta$ ($\bar \rho$).

\begin{figure}[tbp!]
\begin{center}
{\epsfig{figure=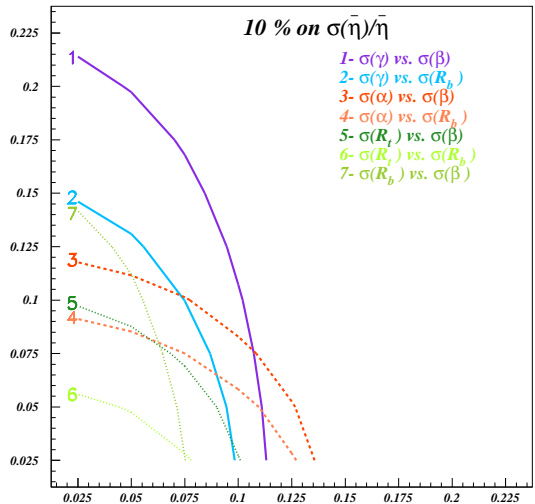,height=8cm}}
\caption[]{\it{ The plot shows the curves of the 10$\%$ relative
precision on $\bar\eta$ as a function of the precision on the variables
of the given strategy.}}
\label{fig:cl_eta}
\end{center}
\end{figure}
\begin{figure}[tbp!]
\begin{center}
{\epsfig{figure=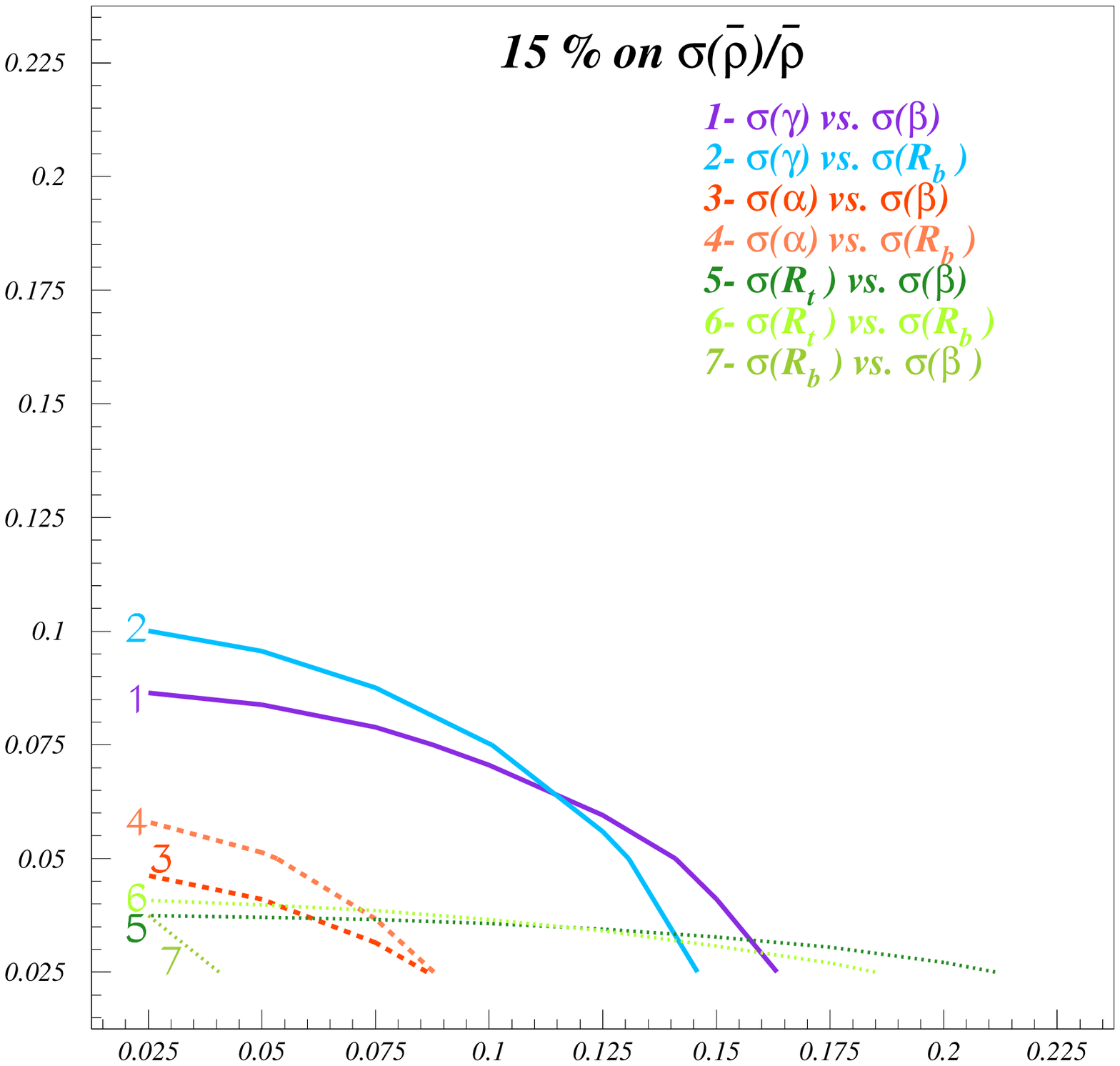,height=8cm}}
\caption[]{\it{The plot shows the curves of the 15$\%$ relative
precision on $\bar\varrho$ as a function of the precision on the variables
of the given strategy.}}
\label{fig:cl_rho}
\end{center}
\end{figure}
\begin{figure}[tbp]
\begin{center}
{\epsfig{figure=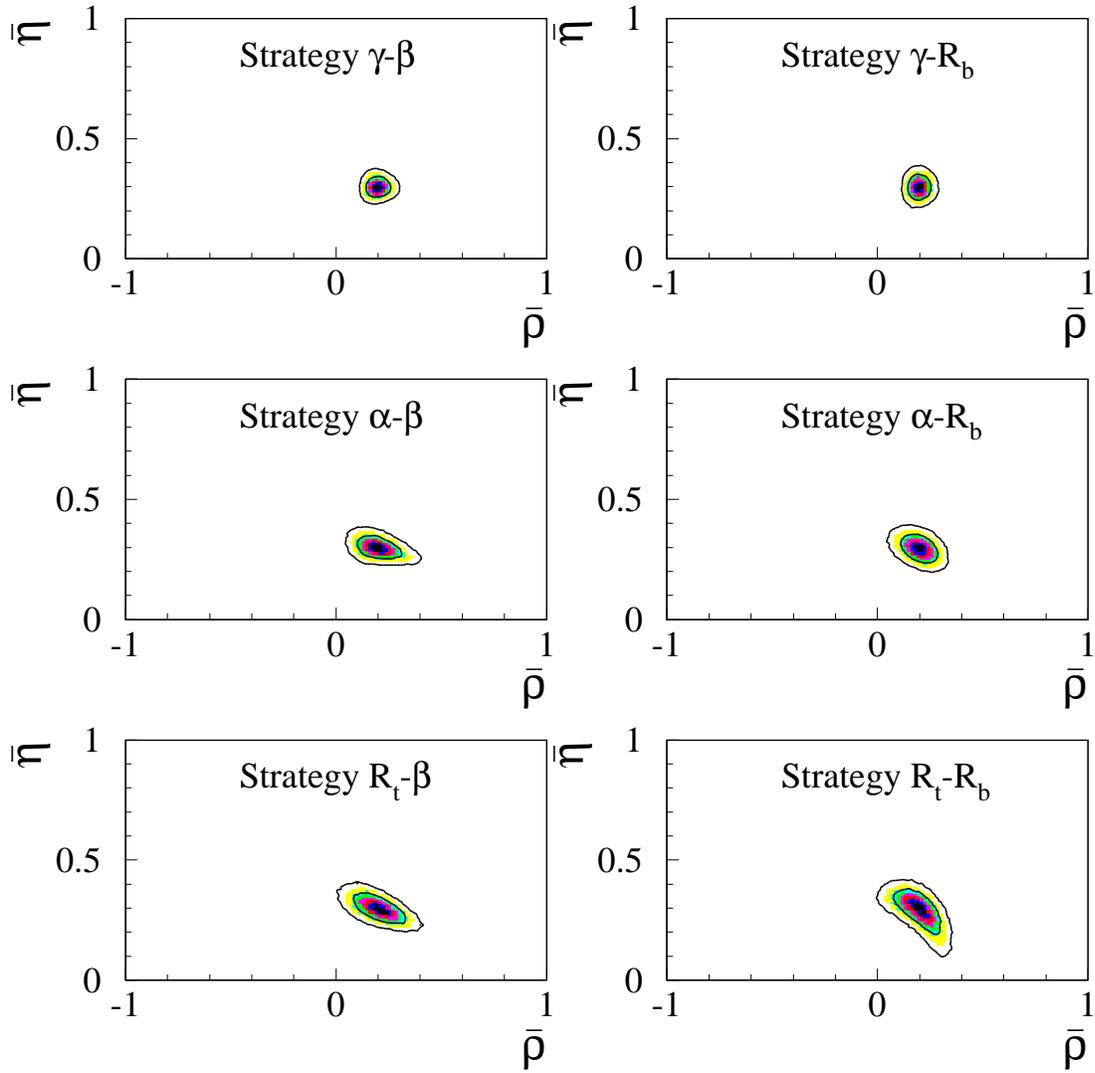,height=16cm}}
\caption[]{\it{The plots show the allowed probability regions 
{(68$\%$ and 95$\%$)} in the $(\bar\varrho,\bar\eta)$ plane
obtained from the leading strategies assuming that each variable is measured
with $10\%$ accuracy.}}
\label{fig:6plots}
\end{center}
\end{figure}
\begin{figure}[tbp]
\begin{center}
{\epsfig{figure=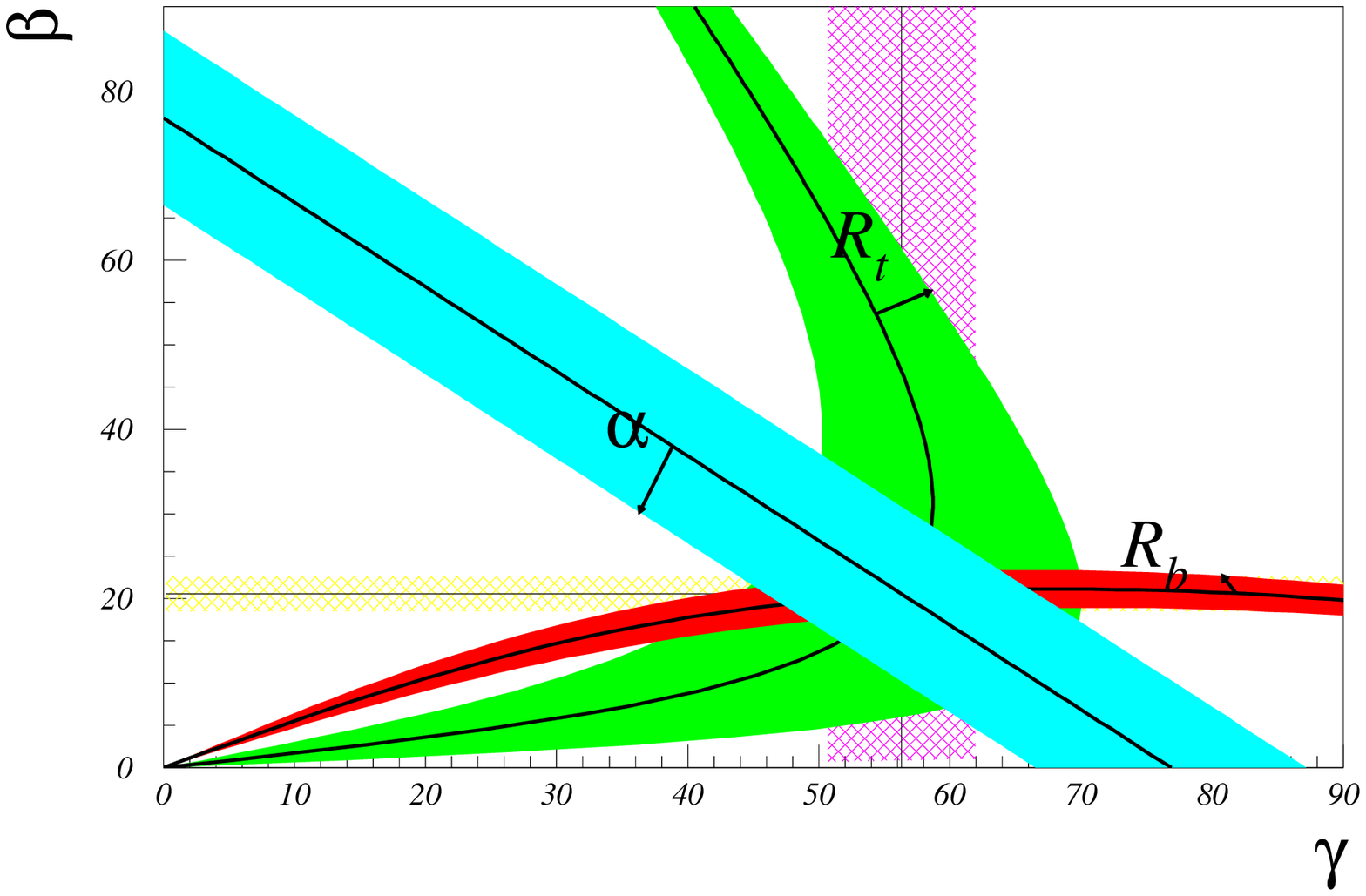,height=5.5cm}}
{\epsfig{figure=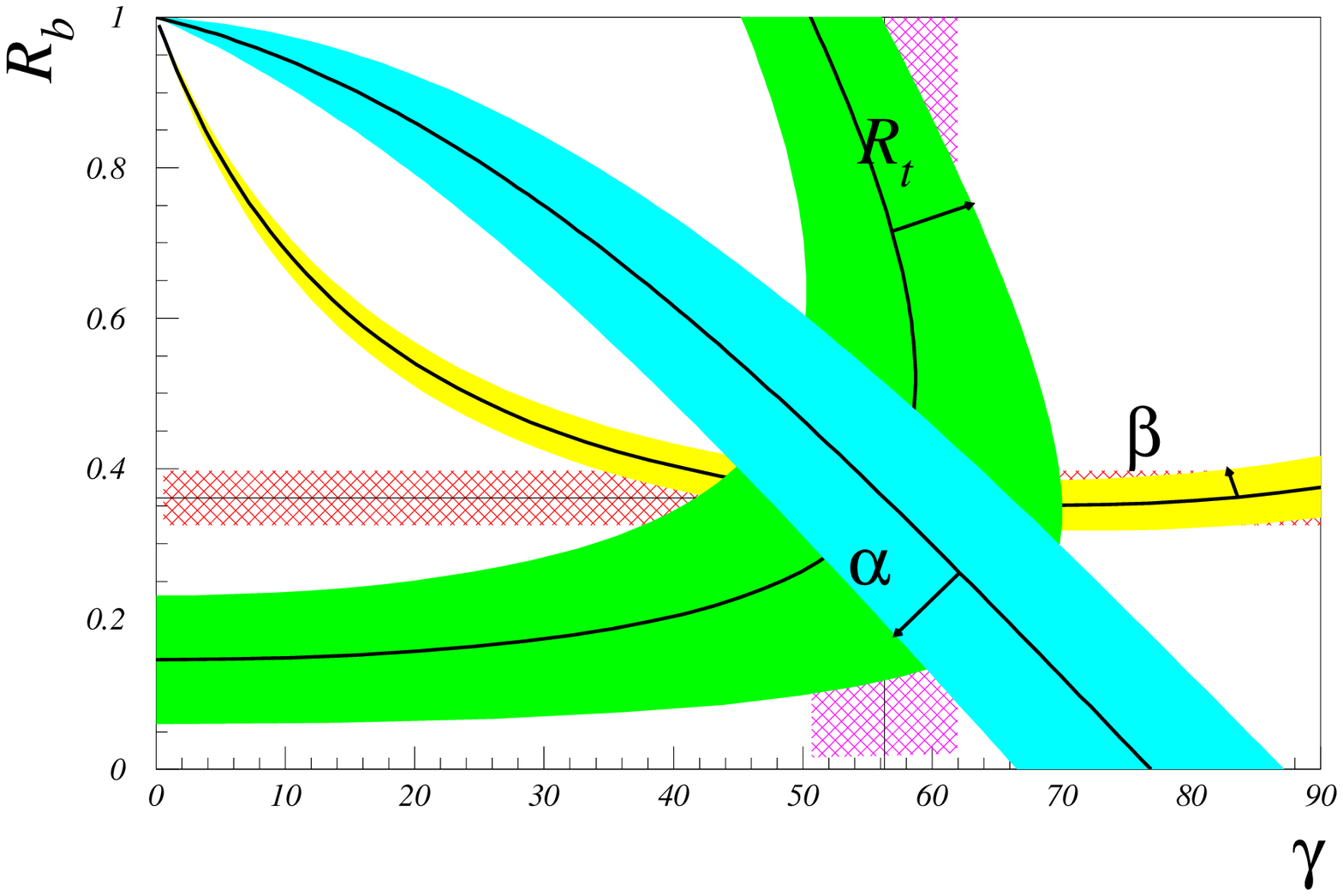  ,height=5.5cm}}\\
{\epsfig{figure=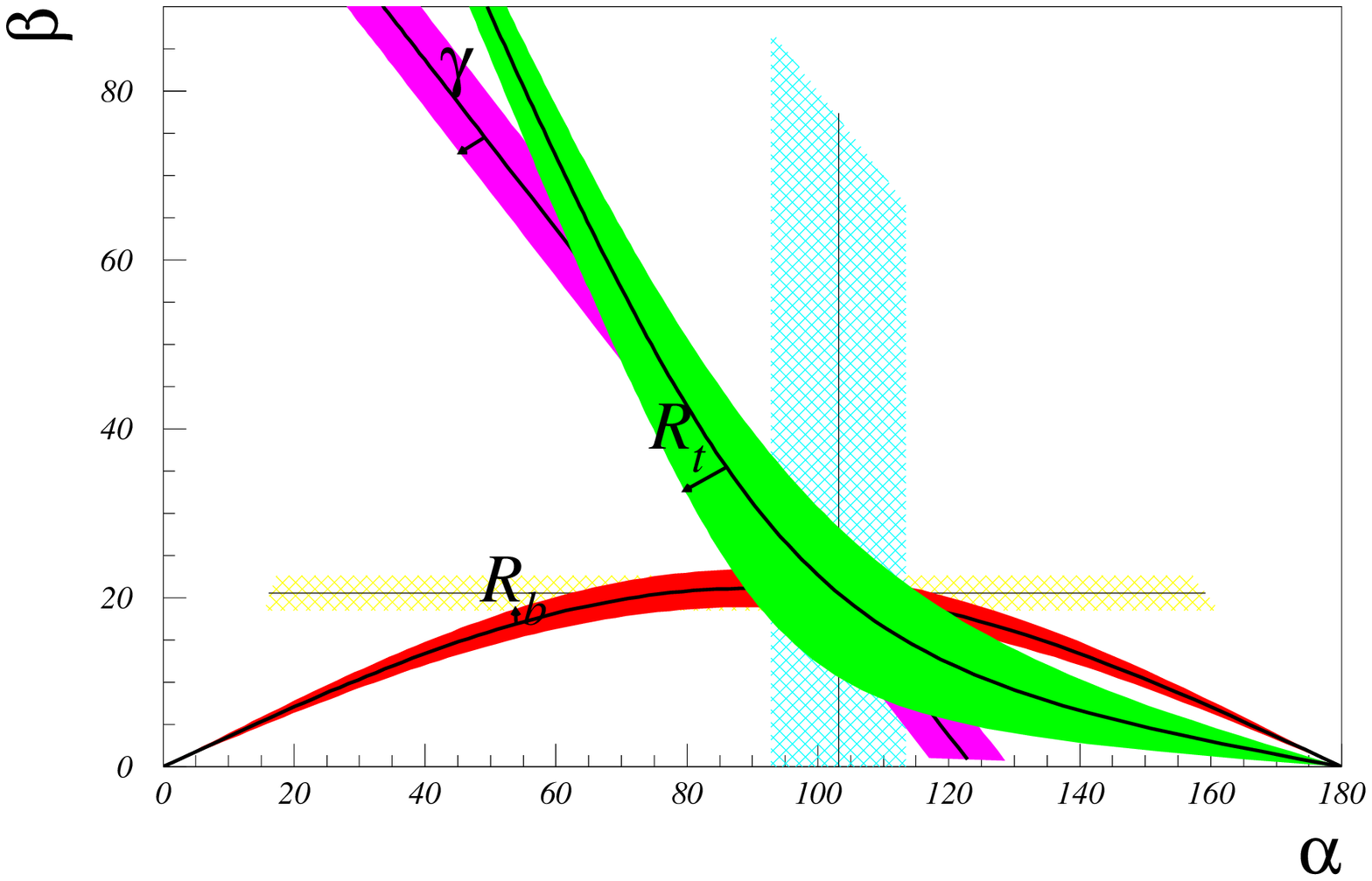,height=5.5cm}}
{\epsfig{figure=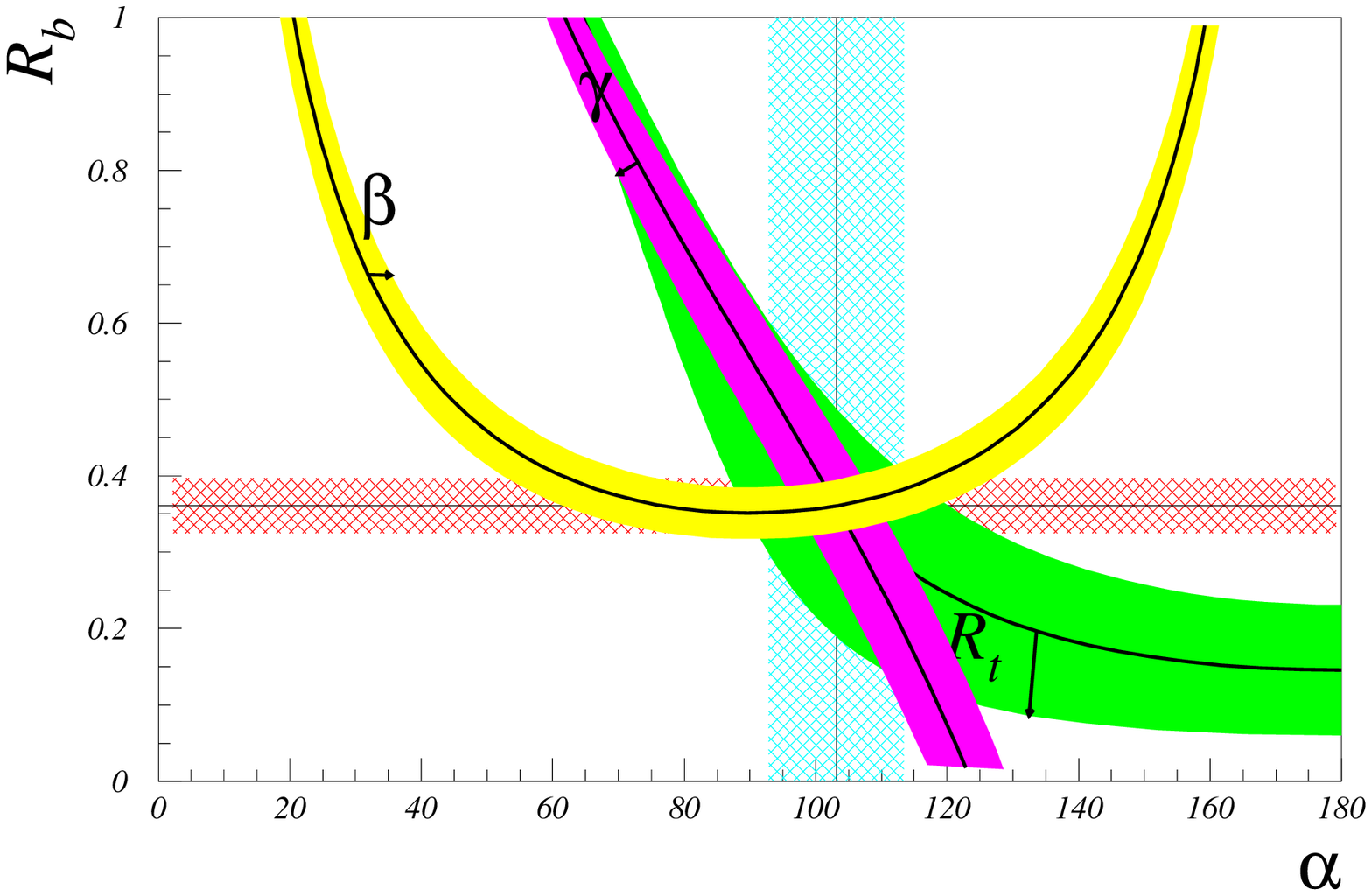  ,height=5.5cm}}\\
{\epsfig{figure=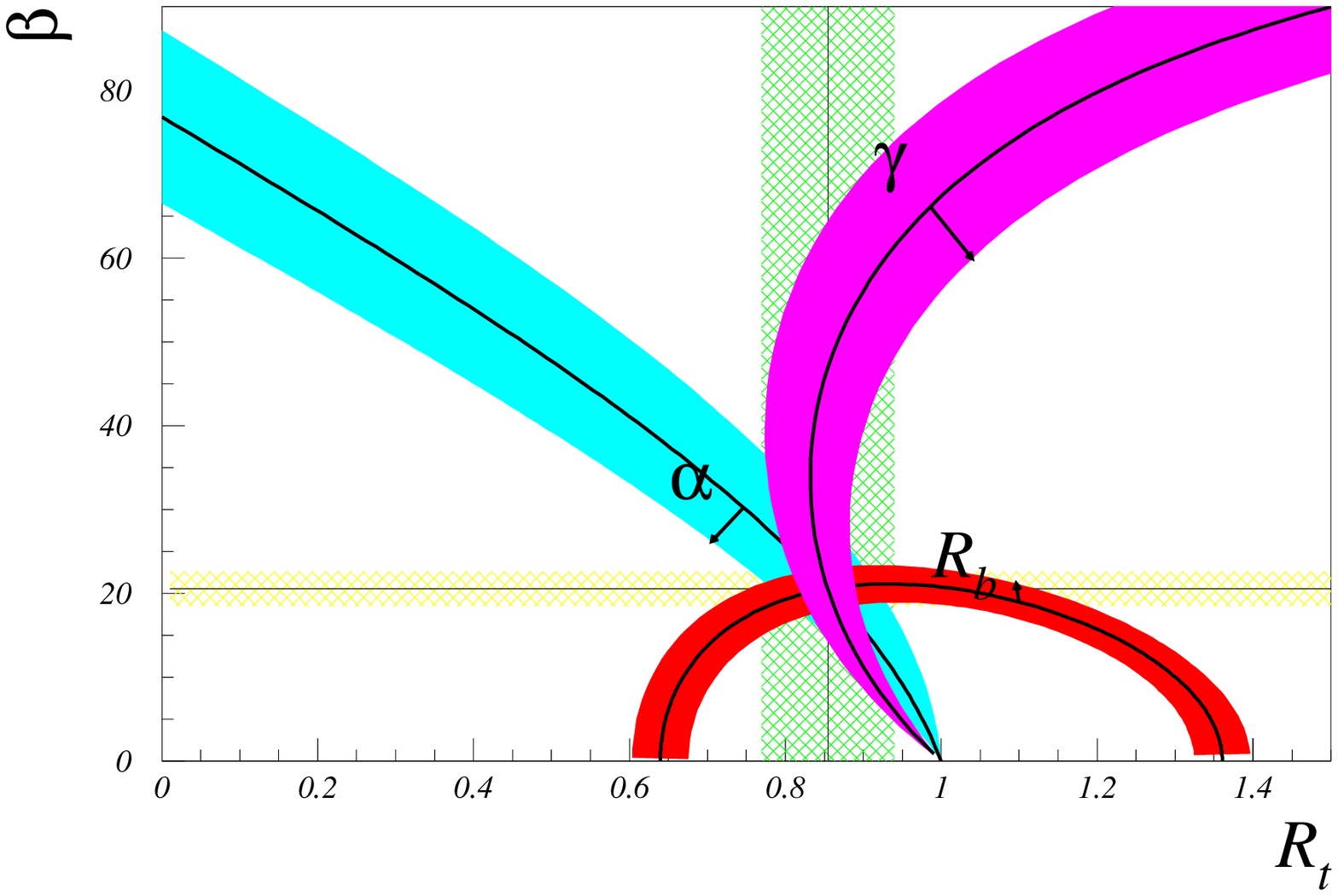   ,height=5.5cm}}
{\epsfig{figure=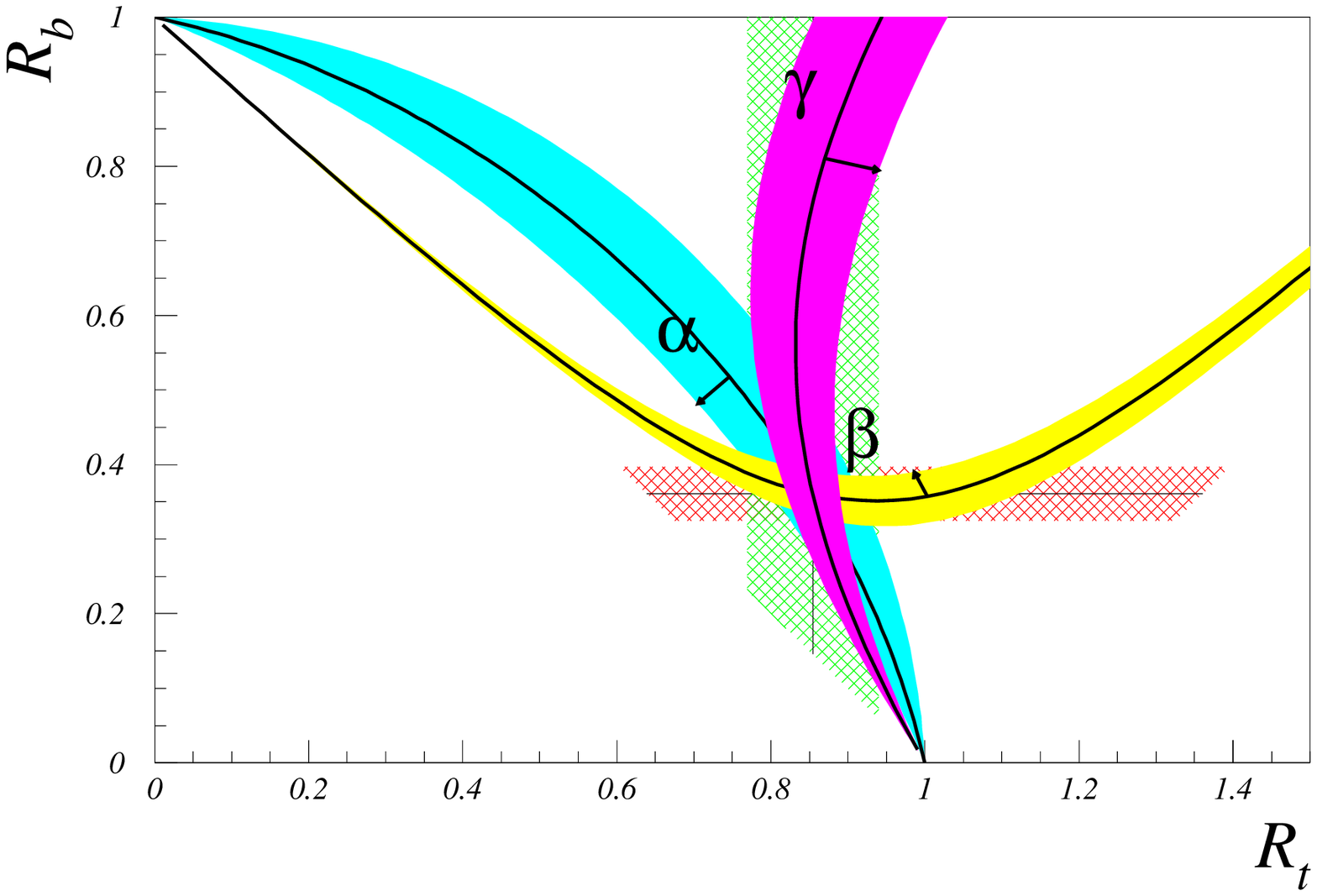     ,height=5.5cm}}
\caption[]{\it{The plots show the different constraints (assuming a relative error
of 10$\%$) in the different planes corresponding to the leading stategies of
equation \ref{ranking1}. The small arrow indicates the range corresponding to an 
increase of 10$\%$ of the corresponding quantity.}}
\label{fig:stra}
\end{center}
\end{figure}

The main results of this analysis are depicted in 
figures \ref{fig:cl_eta}, \ref{fig:cl_rho}, \ref{fig:6plots} and \ref{fig:stra}. 
In figures \ref{fig:cl_eta} and \ref{fig:cl_rho} we plot the 
correlation between the precisions on the variables relevant for a given 
strategy required to reach the assumed precision on $\bar\eta$ and
$\bar\varrho$, respectively. For this exercise we have used, for
$\bar \eta$ and $\bar \rho$, the central values obtained using the input of 
table \ref{inputs}.

 Obviously strategies described by curves in
figures \ref{fig:cl_eta} and \ref{fig:cl_rho} that lie far from the origin 
are more effective in the determination of the unitarity  triangle than
those corresponding to curves placed close to the origin.

Figures \ref{fig:cl_eta} and \ref{fig:cl_rho}
reveal certain hierarchies within the strategies in question. In order to 
find these hierarchies and to eliminate the weakest ones not shown in these 
figures we divided first the five variables under consideration 
into two groups:
\be\label{group}
(R_t,\alpha,\gamma),\qquad (R_b,\beta)~.
\ee

It turned out then that the four strategies $(R_t,\alpha)$, $(R_t,\gamma)$,
$(\alpha,\gamma)$ and $(R_b,\beta)$ which involve pairs of variables
belonging to the same group are not particularly useful in the determination
of $(\bar\varrho,\bar\eta)$.  In the case of $(R_b,\beta)$ this is related to
the existence of two possible solutions as stated above. If one of these 
solutions can easily be excluded on the basis of $R_t$, then the 
effectiveness of this strategy can be increased. 
We have therefore included this strategy in our 
numerical analysis. The strategy $(R_t,\gamma)$ turns out to be less
useful in this respect. Similarly the strategy $(\gamma,\alpha)$ is
not particularly useful due to strong correlation between the variables
in question as discussed previously by many authors in the literature.

The remaining six strategies that involve pairs of variables belonging 
to different groups in (\ref{group}) are all interesting.
While some strategies are better suited for the determination of $\bar\eta$ 
and the other for $\bar\varrho$, as clearly seen in 
figures \ref{fig:cl_eta} and \ref{fig:cl_rho}, 
on the whole a clear ranking of
strategies seems to emerge from our analysis.

If we assume the same relative error on $\alpha$, $\beta$, $\gamma$,
$R_b$ and $R_t$ we find the following hierarchy:
\be\label{ranking1}
1)~(\gamma,\beta), \quad (\gamma,R_b) \qquad 
2)~(\alpha,\beta), \quad (\alpha,R_b) \qquad 
3)~(R_t,\beta), \quad (R_t,R_b), \quad (R_b,\beta).
\ee

We observe that in particular the strategies involving $R_b$ and $\gamma$ are
very high on this ranking list. This is related to the fact that 
$R_b<0.5<R_t$ and consequently the action in the $(\bar\varrho,\bar\eta)$
plane takes place closer to the origin of this plane than to the corner 
of the UT involving the angle $\beta$. Consequently the accuracy on 
$R_b$ and $\gamma$ does not have to be as high as for $R_t$ and $\beta$
in order to obtain the same accuracy for $(\bar\varrho,\bar\eta)$.
This is clearly seen in figures \ref{fig:cl_eta} and \ref{fig:cl_rho}.

This analysis shows how important is the determination of $R_b$ and $\gamma$
in addition to $\beta$ that is already well known.
On the other hand the strategy involving $R_t$ and $\beta$ will be 
most probably the cleanest one before the LHC experiments  unless
the error on $\gamma$ from B-factories and Tevatron can be  
decreased to $10^\circ$ and $R_b$ is significantly better known.

The strategies involving $\alpha$ are in our second best class. 
However, it has to be noticed that in order to get 10$\%$(15$\%$) relative 
precision on $\bar\eta$($\bar\rho$) it is necessary 
(see figures \ref{fig:cl_eta} and \ref{fig:cl_rho}) 
to determine $\alpha$ with better than 10$\%$ relative precision.
 If $\sin 2\alpha$ could be directly measured this could be soon achieved
due to its high sensitivity  to $\alpha$ \cite{BBSIN,BBNS2}.
However, from the present perspective a direct measurement of $\sin 2\alpha$ 
appears to be very difficult in view
of the penguin pollution that could be substantial in view of the most 
recent data from Belle \cite{Bellealpha}. On the other hand, 
as the BaBar data 
\cite{BaBaralpha} do not signal 
this pollution, it may eventually turn out that a useful direct 
measurement of $\sin 2\alpha$ can be soon achieved. The most recent 
theoretical discussions can be found in \cite{FLMA02}.

We have also performed a numerical analysis for the strategies in which
$|\bar \eta|$ can be directly measured. The relevant formulae are given in 
(\ref{S11}) and (\ref{S12}). It turns out that the strategy 
$(\gamma,\bar\eta)$ can be put in the 
first best class in (\ref{ranking1}) together with the strategies 
$(\gamma,\beta)$ and $(\gamma,R_b)$.

In figure \ref{fig:6plots} we show the resulting regions in the 
$(\bar\varrho,\bar\eta)$
plane obtained from leading strategies assuming that each variable is measured
with $10\%$ accuracy. This figure is complementary to 
figures \ref{fig:cl_eta} and \ref{fig:cl_rho} and demonstrates clearly the 
ranking given in (\ref{ranking1}).

While at present the set (\ref{I2}) appears to be the leading candidate for 
the fundamental parameter set in the quark flavour physics for the coming 
years, it is not clear which set will be most convenient in the second half 
of this decade when the B-factories and Tevatron will improve considerably 
their measurements and LHC will start its operation. Therefore it is of 
interest to investigate how the measurements of three variables out of 
$\alpha,~\beta,~\gamma~,R_b$ and $R_t$ will determine the allowed values for 
the remaining two variables. We illustrate this in figure \ref{fig:stra} assuming 
a relative error of $10\%$ for the constraints used in each plot. While this 
figure is self explanatory a striking feature consistent with the hierarchial 
structure in (\ref{ranking1}) can be observed. While the measurements of 
$(\alpha,R_t,R_b)$ and $(\alpha,\beta,R_t)$ as seen in the first two 
plots do not appreciably constrain 
the parameters of the two leading strategies $(\beta,\gamma)$ 
and $(R_b,\gamma)$, respectively, the opposite is true in the last two plots. 
There the measurements of $(R_b,\gamma,\alpha)$ and $(\beta,\gamma,\alpha)$ 
give strong constraints in the $(\beta,R_t)$ and $(R_b,R_t)$ plain, 
respectively.  

The last two plots illustrate also clearly that 
measuring only $\alpha$ and $\gamma$ does not provide a strong constraint on 
the unitarity triangle.

 Finally we would like to emphasize that our ranking does not take into 
account the fact that the assumed precision on some variables can be easier 
to achieve than for other variables. While this is difficult to 
incorporate into our analysis at present, it could be in principle possible
when our understanding of theoretical uncertainties in various determinations
improves.

\section{ Presently available strategies}
\label{sec:present}

\begin{table}[htb]
\begin{center}
\footnotesize{
\begin{tabular}{|c|c|c|c|c|}
\hline
 Parameter &  Value & Gaussian &  Uniform      & Ref. \\
           &        & $\sigma$ &  half-width   &      \\
\hline
    $\lambda$               & $0.2210$   &  0.0020    &         -     &  \cite{ref:ckmworkshop} \\
\hline
$\left | V_{cb} \right |$(excl.) & $ 42.1  \times 10^{-3}$  & $ 2.1 \times 10^{-3}$ 
                                 &              -           & \cite{ref:ArtusoBarberio}\\
$\left | V_{cb} \right |$(incl.) & $ 40.4  \times 10^{-3}$  & $ 0.7 \times 10^{-3}$ 
                                 & $ 0.8 \times 10^{-3}$    & \cite{ref:ArtusoBarberio}\\
$\left | V_{cb} \right|$(ave.) & $ 40.6 \times 10^{-3}$ & \multicolumn{2}{|c|}{$ 0.8 \times 10^{-3}$ ${^*}$ }  &  \\
 \hline
$\left | V_{ub} \right |$(excl.) & $ 32.5  \times 10^{-4}$ & $ 2.9 \times 10^{-4}$ 
                                 & $ 5.5 \times 10^{-4}$ & \cite{ref:ckmworkshop}\\
$\left | V_{ub} \right |$(incl.) & $ 40.9  \times 10^{-4}$ & $ 4.6 \times 10^{-4}$ 
                                 & $ 3.6 \times 10^{-4}$ & \cite{ref:ckmworkshop}\\
$\left | V_{ub} \right |$(ave.)  & $ 36.3 \times 10^{-4}$ & \multicolumn{2}{|c|}{$ 3.2 \times 10^{-4}$ ${^*}$  } &\\
 \hline
 $\left | V_{ub} \right |$/$\left | V_{cb} \right|$(ave.)
                                 & 0.089 & \multicolumn{2}{|c|}{0.008${^*}$}  & \\
 \hline
$\Delta M_d$                      & $0.503~\mbox{ps}^{-1}$ & $0.006~\mbox{ps}^{-1}$ 
                                  & -- & \cite{ref:lepbosc}  \\
$\Delta M_s$  & $>$ 14.4 ps$^{-1}$ at 95\% C.L. & \multicolumn{2}{|c|}
{sensitivity 19.2 ps$^{-1}$} & \cite{ref:lepbosc}  \\
$m_t$ & $167~GeV$ & $ 5~GeV$ & -- & \cite{ref:top} \\
$f_{B_d} \sqrt{\hat B_{B_d}}$ & $235~MeV$  & $33~MeV$ &  $^{+0}_{-24}~MeV$  & \cite{ref:lellouch} \\
$\xi=\frac{ f_{B_s}\sqrt{\hat B_{B_s}}}{ f_{B_d}\sqrt{\hat B_{B_d}}}$ 
                                  & 1.18   & 0.04 & $^{+0.12}_{-0.00}$ & \cite{ref:lellouch} \\
 \hline
$\hat B_K$                    & 0.86   & 0.06 & 0.14 & \cite{ref:lellouch} \\
 \hline
         sin 2$\beta$             & 0.734   & 0.054 & - & \cite{ref:sin2b} \\ \hline
\hline
\end{tabular} }
\caption[]{ \small { Values of the relevant quantities entering into the expressions of 
$\vub$, $\Delta M_d$ and $\Delta M_s$. In the third and fourth columns the Gaussian and 
the flat part of the uncertainty are given, respectively. 
Here $m_t\equiv \overline{m}_t(m_t)$. }}

\label{inputs}
\end{center}
\end{table}

\begin{table}[htb!]
\begin{center}
\begin{tabular}{|c|c|c|}
\hline
  Strategy       &        $\bar \rho$            &      $\bar \eta$              \\
  ($R_t,\beta$)  &    0.157 $^{+0.056}_{-0.054}$ &    0.367 $^{+0.036}_{-0.034}$ \\
                 &       [0.047-0.276]           &        [0.298-0.439]          \\
  ($R_t,R_b$)    &    0.161$^{+0.055}_{-0.057}$  &   0.361 $^{+0.041}_{-0.045}$  \\  
                 &       [0.043-0.288]           &        [0.250-0.438]          \\
  ($R_b,\beta$)  &  0.137 $^{+0.135}_{-0.135}$   &   0.373 $^{+0.049}_{-0.063}$  \\   
                 &       [-0.095-0.357]          &        [0.259-0.456]          \\
\hline
\end{tabular}
\caption[]{ \small {Results for $\bar \rho$ and $\bar \eta$ for the three
indicated strategies using the present knowledge summarized in Table 
\ref{inputs}. For the strategy ($R_t,\beta$), the solution compatible with 
the region selected by the $R_b$ constraint has been considered. In squared 
brackets the 95$\%$ probability regions are also given.}}
\label{results}
\end{center}
\end{table}

\begin{figure}[htb!]
\begin{center}
{\epsfig{figure= 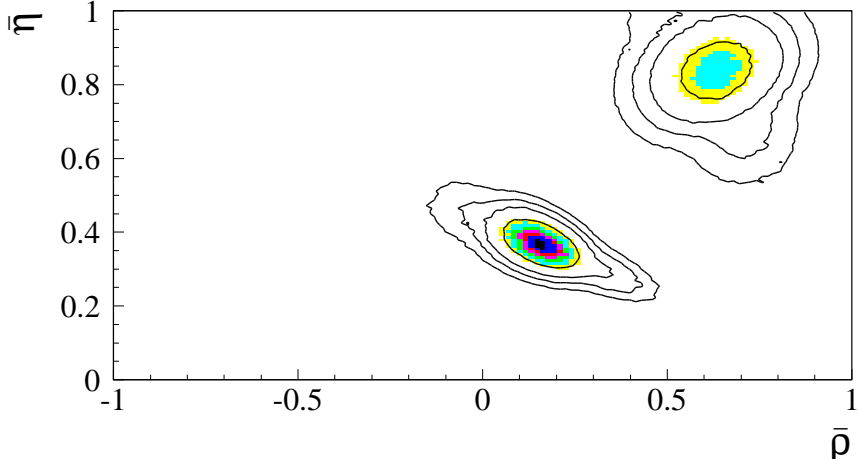,height=7cm}}
\caption[]{\it{The plot shows the presently allowed probability
regions {(68$\%$,95$\%$,99$\%$ and 99.9$\%$)} in the 
($\bar\rho,\bar\eta$) plane using the $(R_t,\beta)$ strategy: the 
direct measurement of sin 2$\beta$ and $R_t$ from $\Delta M_d$ 
and ${\Delta M_d}/{\Delta M_s}$}}
\label{fig:rtbeta_real}
\end{center}
\end{figure}

\begin{figure}[htb!]
\begin{center}
{\epsfig{figure= 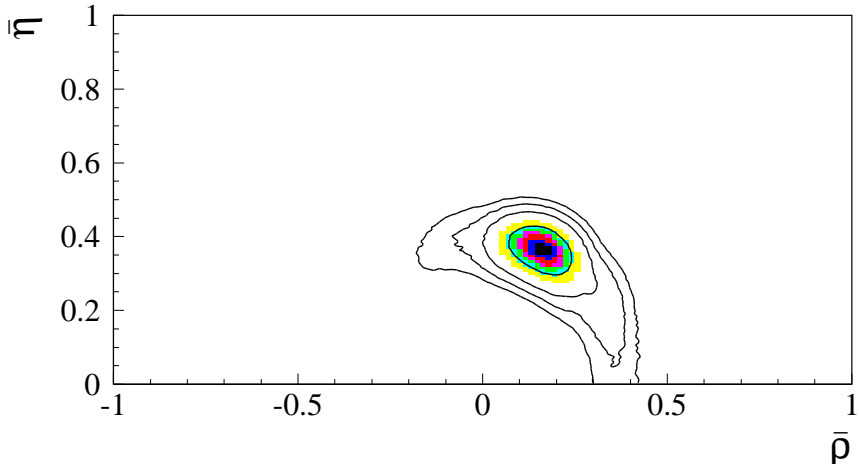,height=7cm}}
\caption[]{\it{The plot shows the allowed probability 
regions {(68$\%$,95$\%$,99$\%$ and 99.9$\%$)} in the 
($\bar \rho,\bar \eta$) plane using the $(R_t,R_b)$ strategy: 
$R_t$ from $\Delta M_d$ and ${\Delta M_d}/{\Delta M_s}$ and $R_b$ from 
$\vub$}}
\label{fig:rtrb_real}
\end{center}
\end{figure}

\begin{figure}[htb!]
\begin{center}
{\epsfig{figure= 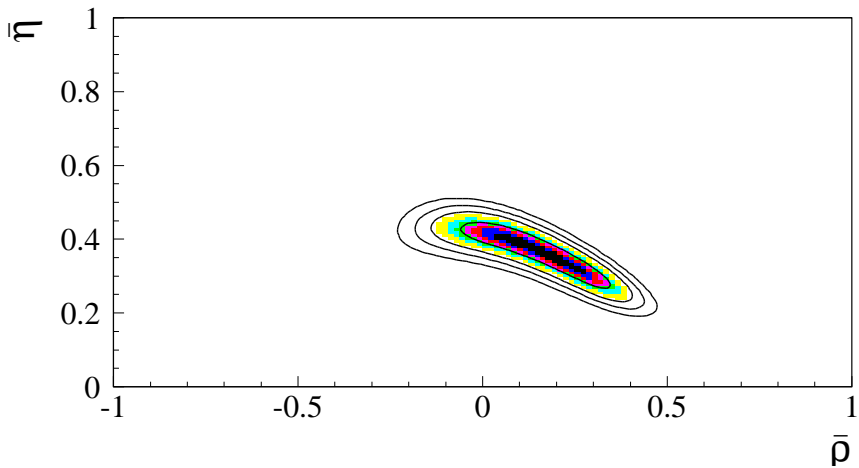,height=7cm}}
\caption[]{\it{The plot shows the allowed probability 
regions {(68$\%$,95$\%$,99$\%$ 
and 99.9$\%$)} in the 
($\bar \rho,\bar \eta$) plane using the $(R_b,\beta)$ strategy: 
direct measurement of sin 2$\beta$ and $R_b$ from 
$\left|{V_{ub}}/{V_{cb}}\right|$.}}
\label{fig:rbbeta_real}
\end{center}
\end{figure}

\begin{figure}[htb!]
\begin{center}
{\epsfig{figure=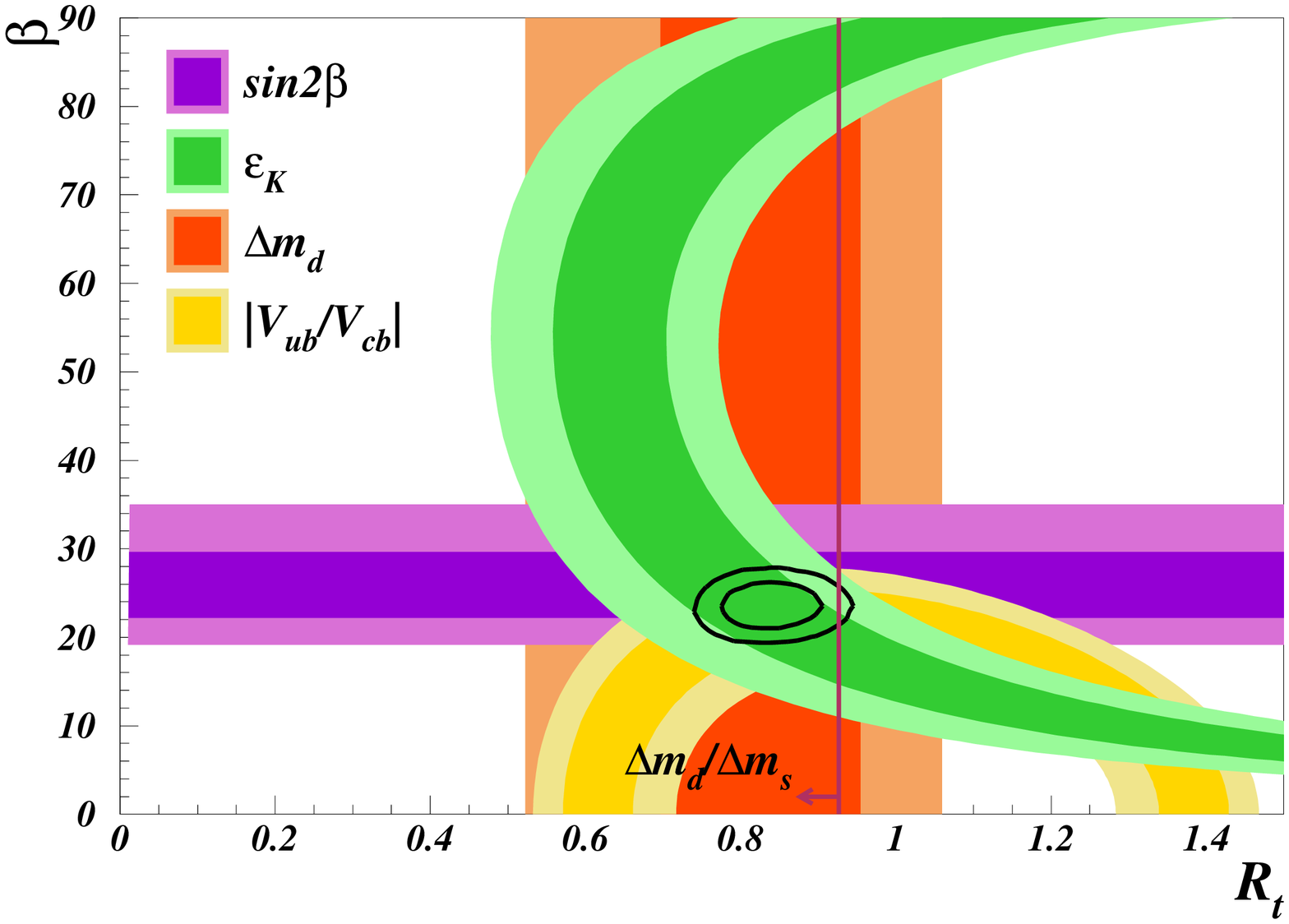,height=10cm}}
\caption[]{\it{The plot shows the allowed probability 
regions {(68$\%$ and 95$\%$)} in the ($R_t,\beta$) plane.
Different constraints are also shown. The line shown for 
${\Delta M_d}/{\Delta M_s}$ constraint corresponds to the 95$\%$ C.L. 
limit on ${\Delta M_s}$ }}
\label{fig:rtbeta}
\end{center}
\end{figure}

At present the concrete results can be obtained only for the strategies
($R_t,\beta$), ($R_b,R_t$) and ($R_b,\beta$)  as no direct measurements of
$\gamma$ and $\alpha$ are available. The most recent discussions of 
the strategies for the determination of $\alpha$ and $\gamma$  
with the relevant references can be found in \cite{BBNS2,FLMA02} and 
\cite{FLgamma}.

The results for $\bar \rho$ and $\bar \eta$ for the three strategies in
question are presented in table \ref{results} and in figures 
\ref{fig:rtbeta_real}, \ref{fig:rtrb_real} and \ref{fig:rbbeta_real}. 
To obtain these results we have used the direct measurement of 
sin 2$\beta$ \cite{ref:sin2b}, $R_t$ as extracted from
$\Delta M_d$ and ${\Delta M_d}/{\Delta M_s}$ by means of the formulae in 
\cite{Erice,ref:haricot} and $R_b$ as extracted from 
$\vub$.
The experimental and theoretical inputs are summarized in table \ref{inputs}
and the methods are described in \cite{ref:haricot}.
The errors with stars in table \ref{inputs} are the r.m.s. of 
the distributions, resulting upon the convolution of the two different 
determinations, and thus indicative. In the fit full distributions have 
been used.

It should be emphasized that these three presently available strategies are 
the weakest among the leading strategies listed in (\ref{ranking1}). 
Among them $(R_t,\beta)$ and $(R_t,R_b)$ appear to be superior to 
$(R_b,\beta)$ at present. We expect that once $\Delta M_s$ has been measured 
and the error on $\sin 2\beta$ reduced, the strategy $(R_t,\beta)$ will be 
leading among these three. Therefore in figure \ref{fig:rtbeta} we show how 
the presently available 
constraints look like in the $(R_t,\beta)$ plane.

\section{An Update on Minimal Flavour Violation Models}
\label{sec:mfvm}
The simplest class of extentions of the SM are the models 
in which only the SM operators in the effective weak Hamiltonian are 
relevant  and in which flavour 
violation is entirely governed by the CKM matrix. In these models CP 
violation is governed then solely by the KM phase. The Two-Higgs-Doublet 
model and the MSSM at low $\tan\beta=v_2/v_1$ belong to this class of
models. 
 We will call this scenario ``Minimal Flavour 
Violation" (MFV) \cite{UUT} being aware of the fact that for some authors 
MFV means a more general framework in which also new operators can give 
significant contributions. 
See for instance the recent discussions in 
\cite{BOEWKRUR,AMGIISST}.

The unitarity triangle in specific supersymmetric models of the MFV-type has
been extensively analyzed in \cite{ALILOND} and general properties of the 
MFV models have been pointed out in 
\cite{BLO,UUT,BF01,BCRS1,ABRB,Perez,NIR2002}.
 The interesting virtue of the MFV models is
that with respect to $B^0_{d,s}-\bar B^0_{d,s}$ mixings and the CP-violating
parameter $\varepsilon_K$, they all can be parametrized by a single function 
$F_{tt}$ \cite{MIPORO,ALILOND}. 
In the SM, $F_{tt}$ results from box diagrams with top quark and 
$W^\pm$ exchanges with $F_{tt}=2.39\pm 0.12$. Beyond the SM, $F_{tt}$ depends 
on various new parameters like the masses of charginos, squarks and charged 
Higgs particles and it can in principle take any value.

One of the interesting properties of the MFV models is the existence of 
the universal unitarity triangle (UUT) \cite{UUT} that can be constructed 
from quantities in which all the dependence on new physics cancels out
or is negligible like in tree level decays from which $|V_{ub}|$ and 
$|V_{cb}|$ are extracted. The values of $\bar\varrho$, $\bar\eta$, $\alpha$,
$\beta$, $\gamma$, $R_b$, and $R_t$  resulting from 
this determination are
the ``true" values that are universal within the MFV models. 
Various strategies for the determination of the UUT are discussed 
in \cite{UUT}.

The presently available quantities that do not depend on the new physics 
parameters within the MFV-models and therefore can be used to determine 
the UUT are $R_t$ from $\Delta M_d/\Delta M_s$ by means of
\be\label{RTA}
R_t=\frac{\xi}{\lambda} \sqrt{\frac{\Delta M_d}{\Delta M_s}}
\ee
with $\xi$ defined in table~\ref{inputs},
$R_b$ from $\vub$ by means of (\ref{2.94})   and $\sin 2\beta$ 
extracted from the CP asymmetry in $B^0_d\to \psi K_S$. 
On the other hand $\varepsilon_K$ and $R_t$ from $\Delta M_d$ alone cannot 
be used in this construction as they both depend explicitly on $F_{tt}$.
Formula (\ref{RTA}) is an excellent approximation.

\begin{figure}[htb!]
\begin{center}
{\epsfig{figure=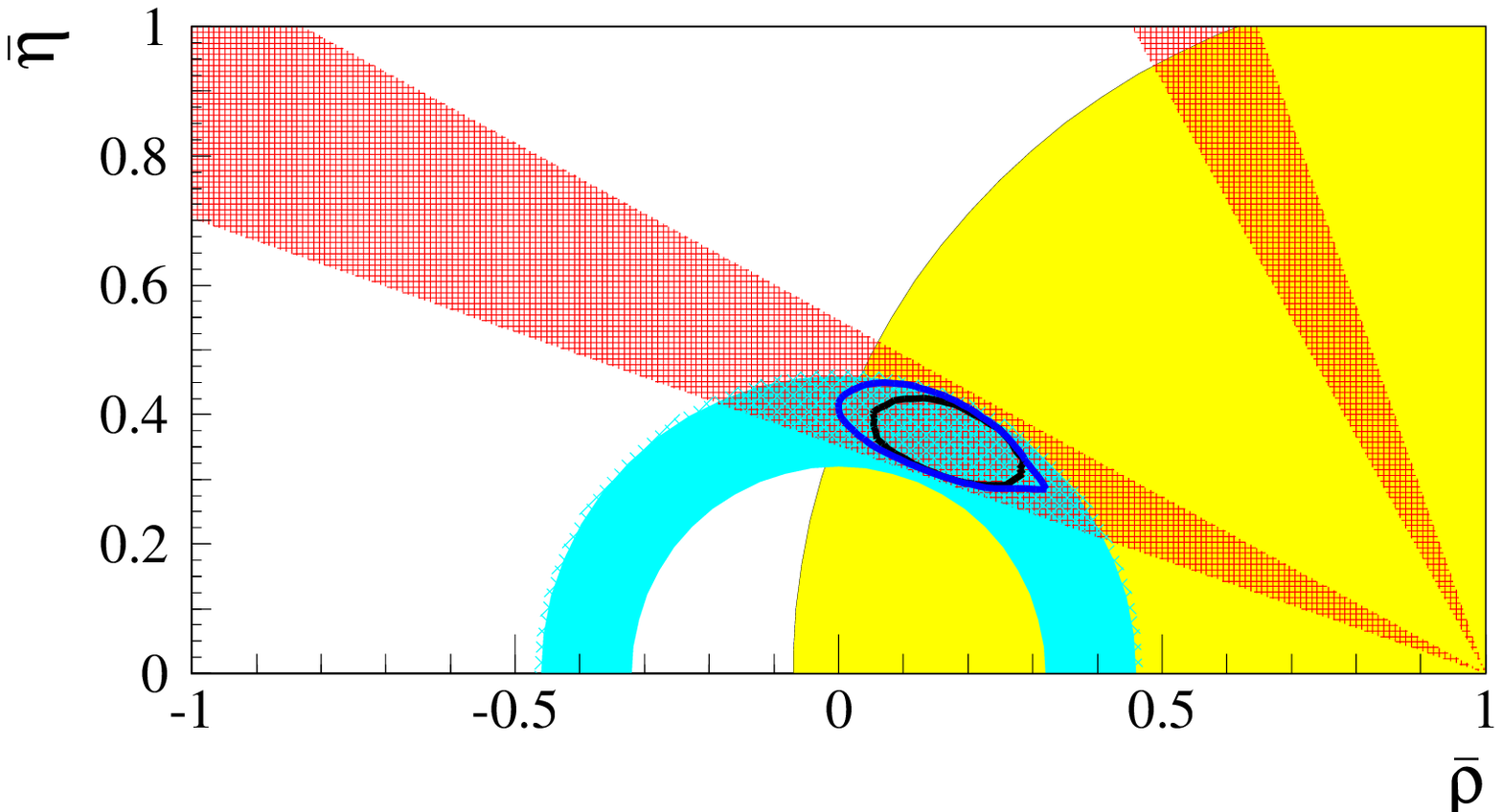,height=9cm}}
\caption[]{\it{The plot shows the allowed 95$\%$ probability region in the 
$(\bar\varrho,\bar\eta)$ plane,
consistent with UUT; the individual 95$\%$ regions for the constraint from 
$sin 2 \beta$, $\Delta M_s$ and $R_b$ are also shown. The narrower region 
corresponds to the allowed 95$\%$ probability 
region consistent with SM. The results are
obtained using the fit procedure described in \cite{ref:haricot}}}
\label{fig:figmfv}
\end{center}
\end{figure}

Using the necessary input of table \ref{inputs} that includes the most recent data 
on $\sin 2\beta$ from BaBar and Belle \cite{ref:sin2b}, we find the allowed 
universal 
region for $(\bar\varrho,\bar\eta)$ in the MFV models shown in 
figure \ref{fig:figmfv}. 
Similar analysis has been done in \cite{AMGIISST}.
The central values, the errors and 
the $95\%$ (and $99\%$) probability ranges for various quantities of 
interest related to this UUT are collected in table \ref{mfv}.
These quantities are compared with the corresponding results in the SM as shown 
in figure \ref{fig:figmfv} and in table \ref{mfv}.
To this end all available constraints, 
that is also $\varepsilon_K$ and $\Delta M_d$ have been used, following the
procedure described in \cite{ref:haricot}.
Other recent analyses of the UT in the SM can be found in 
\cite{ALILOND,Lacker,ATSO,bologna,talks}.

\begin{table}[htb!]
\begin{center}
\begin{tabular}{|c|c|c|}
\hline
  Strategy       &               UUT              &            SM                \\
  $\bar {\eta}$  &         0.369 $\pm$ 0.032      &     0.357 $\pm$ 0.027        \\ 
                 &   (0.298-0.430) [0.260-0.449]  & (0.305-0.411)  [0.288-0.427] \\
  $\bar {\rho}$  &         0.151 $\pm$ 0.057      &     0.173 $\pm$ 0.046        \\
                 &   (0.034-0.277) [-0.023-0.358]  & (0.076-0.260)  [0.043-0.291] \\
  $\sin 2\beta$  &   0.725 $^{+0.038}_{-0.028}$   &  0.725 $^{+0.035}_{-0.031}$  \\
                 &   (0.661-0.792) [0.637-0.809]  & (0.660-0.789)  [0.637-0.807] \\
  $\sin 2\alpha$ &        0.05 $\pm$ 0.31         &   -0.09 $\pm$ 0.25           \\
                 &   (-0.62-0.60)  [-0.89-0.78]   & (-0.54-0.40)   [-0.67-0.54]  \\
  $\gamma$       &         67.5 $\pm$ 9.0         &    63.5 $\pm$ 7.0            \\
  (degrees)      &   (48.2-85.3)   [36.5-93.3]    & (51.0-79.0)    [46.4-83.8]   \\
  $R_b$          &         0.404 $\pm$ 0.023      &   0.400 $\pm$ 0.022          \\
                 &   (0.359-0.450) [0.345-0.463]  & (0.357-0.443)  [0.343-0.457] \\
  $R_t$          &         0.927 $\pm$ 0.061      &   0.900 $\pm$ 0.050          \\
                 &   (0.806-1.048) [0.767-1.086]  & (0.802-0.998)  [0.771-1.029] \\
  $\Delta M_s$   &         17.3$^{+2.2}_{-1.3}$   &   18.0$^{+1.7}_{-1.5}$       \\
  ($ps^{-1}$)    &   (15.0-23.0)   [11.9-31.9]    & (15.4-21.7)    [14.8-25.9]   \\
$\vtd~(10^{-3})$ &          8.36 $\pm$ 0.55       &        8.15 $\pm$ 0.41       \\
                 &   (7.14-9.50)   [6.27-10.00]    & (7.34-8.97)    [7.08-9.22]   \\
  $\vtd/\vts$    &          0.209 $\pm$ 0.014     &      0.205 $\pm$ 0.011       \\
                 &   (0.179-0.238) [0.157-0.252]  & (0.184-0.227)  [0.177-0.233] \\
  $Im \lambda_t$ &         13.5 $\pm$ 1.2         &    13.04 $\pm$ 0.94          \\
    ($10^{-5}$)  &   (10.9-15.9)   [9.4-16.6]     & (11.2-14.9)    [10.6-15.5]   \\
\hline
\end{tabular}
\caption[]{ \small { Values and errors for different quantities using the present 
knowledge summarized in Table \ref{inputs} for the UUT and the SM Triangle,
following the procedure described in \cite{ref:haricot}.
In brackets the 95$\%$ and 99$\%$ probability regions are also given.}}
\label{mfv}
\end{center}
\end{table}

We would like to remark that the measurement $\sin 2\beta=0.734\pm 0.054$ 
implies two solutions for $\beta$ with $\beta\approx (23.6\pm 2.2)^\circ$ and 
$\beta\approx (66.4\pm 2.2)^\circ$. These are shown in fig. \ref{fig:figmfv}.
In doing this we tacitly assume that the function $F_{tt}$ is positive. 
Otherwise as discussed in \cite{BF01} also two solutions with a negative 
$\beta$ would be possible. 
As in the SM and in the MFV models there are no new complex phases
beyond the CKM phase, the measured $\beta$ is the ``true" $\beta$ and not
the sum of a true $\beta$ and some additional phase that could be the 
case in models with additional sources of flavour and CP violation. 
Consequently by imposing the constraint from $R_b$ the solution with the
larger $\beta$ can be eliminated for all MFV models resulting in the 
unique solution with lower $\beta$ as presented in
figure \ref{fig:figmfv} and table \ref{mfv}.

It should be stressed that any MFV model that is inconsistent with the 
broader allowed region in figure \ref{fig:figmfv} and 
the "UUT" column in table \ref{mfv} is ruled out. 
We observe that there is little room for MFV models that in their predictions 
differ from the SM as the most ranges within the SM and MFV models do not 
significantly differ from each other. 
Among the five observables $R_t$, $R_b$, $\alpha$, $\beta$ and $\gamma$:
\begin{itemize}
\item
The allowed ranges for $R_b$ and $\beta$ in the SM and general MFV models
are essentially indistinguishable from each other.
\item
On the other hand the allowed ranges for $R_t$, $\alpha$ and $\gamma$
in the MFV models are larger than in the SM. 
Moreover $\Delta M_s$ in the MFV models can be slightly larger 
than in the SM.
\end{itemize}

Figure \ref{fig:figmfv} and table \ref{mfv} imply that only 
certain ranges for the parameters specific to a given MFV model are allowed. 
We illustrate this with the function $F_{tt}$.
 Including in 
our analysis of the MFV models, $\varepsilon_K$ and $\Delta M_d$ that 
explicitly depend on $F_{tt}$ \cite{Erice} we find the range
\begin{equation}
1.3 \le F_{tt} \le 3.8  ~(95\%~{\rm probability~region}) 
\qquad [1.2 \le F_{tt} \le 5.1] ~(99\%~{\rm probability~region}) 
\end{equation}
This should be compared with the older range $1.2\le F_{tt}\le 5.7$ 
found in \cite{Perez}. Our results are compatible with those of 
\cite{ALILOND} where specific MFV models have been considered.

The allowed ranges for UUT in figure \ref{fig:figmfv} and table \ref{mfv} 
and the range for $F_{tt}$ will be further constrained in the coming years 
through the improved measurements of $R_b$ and $\sin 2\beta$ and in particular 
of $\Delta M_s$. 

\section{Conclusions}
\label{sec:conclusions}
In this paper we have presented a numerical analysis of the unitarity triangle 
from a different point of view, that emphasizes the role of different 
strategies in the precise determination of the unitarity triangle parameters. 
A complete list of the relevant formulae can be found in Section 3. 
While we have found that the pairs $(\gamma,\beta)$,
$(\gamma,R_b)$ and $(\gamma,\bar\eta)$ are most efficient in determining 
$(\bar\varrho,\bar\eta)$, we expect that the pair $(R_t,\beta)$ 
will play the leading role in the UT fits in the coming years, in particular,
when $\Delta M_s$ will be measured and the theoretical error on $\xi$ 
decreased. For this reason we have proposed 
to plot available constraints on the CKM matrix in the $(R_t,\beta)$ plane. 

It will be interesting to compare in the future the allowed ranges 
for $(\bar\varrho,\bar\eta)$ resulting from different strategies in order 
to see whether they are compatible with each other. Any discrepancies 
will signal the physics beyond the SM. We expect that the strategies 
involving $\gamma$ will play a very important role in this comparison.

For the fundamental set of parameters in the quark flavour physics given in 
(\ref{I2}) we find within the SM
\begin{eqnarray}
\vus=0.221\pm 0.002,~\vcb=(40.4\pm0.8) 10^{-3},~  
R_t=0.90\pm0.05,~\beta=(23.2 \pm 1.4)^\circ \nonumber \quad (7.1)
\end{eqnarray}
where the errors represent one standard deviations and the result  
for $\beta$ corresponds to $\sin 2\beta=0.725\pm 0.033$. 

A complete analysis of the usefulness of a given 
strategy should also include the discussion of its experimental feasibility 
and theoretical cleanness. Extensive studies of these two issues can be found
in \cite{BABAR,LHCB,FERMILAB,BF97,Erice}. Again among various strategies, 
the $(R_t,\beta)$ strategy is 
exceptional as the theoretical uncertainties in the determination of these 
two variables are small and the corresponding experiments are presently 
feasible. In the long run, 
when $\gamma$ will be cleanly measured in $B_s\to DK$ decays at LHC and 
constrained through other decays as reviewed in \cite{FLgamma} we expect that 
the strategy $(\gamma,\beta)$ will take over the leading role.
Eventually the independent direct determinations of the five variables
in question will be crucial for the tests of the SM and its extentions.

We have also determined the universal unitarity triangle for the full class of 
MFV-models as defined in \cite{UUT} and we have compared it with the 
unitarity triangle in the SM.
 We have found that the allowed ranges for various 
parameters related to the unitarity triangle do not significantly differ from
the ones found in the SM. The result can be found in table \ref{mfv} 
and figure \ref{fig:figmfv}. 
The updated $95\%$ probability range for the function 
$F_{tt}$ that parametrizes 
different MFV models reads $1.3\le F_{tt}\le 3.8$ to be compared with the 
corresponding range $ 2.16 \le F_{tt}\le 2.62$ in the 
Standard Model.

\vskip 1.5cm

\noindent
{\bf Acknowledgements}

The work has been supported in part by the German Bundesministerium f\"ur
Bildung und Forschung under the contract 05HT1WOA3 and the DFG Project 
Bu. 706/1-1.

\vfill\eject

\end{document}